\documentclass[twocolumn]{bmcart}

\usepackage{amsthm,amsmath}
\RequirePackage[numbers]{natbib}
\usepackage[utf8]{inputenc} 
\usepackage{graphicx}
\usepackage{booktabs}
\usepackage{multirow}
\usepackage{makecell}
\usepackage{textcomp}
\usepackage{stfloats}
\RequirePackage{hyperref}

\startlocaldefs
\endlocaldefs

\begin{document}

\begin{frontmatter}

\begin{fmbox}
\dochead{Research}


\title{Multi-encoder attention-based architectures for sound recognition with partial visual assistance}


\author[
  addressref={aff1},                   
  corref={aff1},                       
  email={wim.boes@esat.kuleuven.be}   
]{\inits{W.B.}\fnm{Wim} \snm{Boes}}
\author[
  addressref={aff1},
  email={hugo.vanhamme@esat.kuleuven.be}
]{\inits{H.V.h.}\fnm{Hugo} \snm{Van hamme}}


\address[id=aff1]{
  \orgdiv{ESAT},             
  \orgname{KU Leuven},          
  \city{Leuven},                              
  \cny{Belgium}                                    
}





\begin{abstractbox}

\begin{abstract} 
Large-scale sound recognition data sets typically consist of acoustic recordings obtained from multimedia libraries. As a consequence, modalities other than audio can often be exploited to improve the outputs of models designed for associated tasks. Frequently, however, not all contents are available for all samples of such a collection: For example, the original material may have been removed from the source platform at some point, and therefore, non-auditory features can no longer be acquired. 

We demonstrate that a multi-encoder framework can be employed to deal with this issue by applying this method to attention-based deep learning systems, which are currently part of the state of the art in the domain of sound recognition. More specifically, we show that the proposed model extension can successfully be utilized to incorporate partially available visual information into the operational procedures of such networks, which normally only use auditory features during training and inference. Experimentally, we verify that the considered approach leads to improved predictions in a number of evaluation scenarios pertaining to audio tagging and sound event detection. Additionally, we scrutinize some properties and limitations of the presented technique.
\end{abstract}


\begin{keyword}
\kwd{multi-encoder}
\kwd{transformer}
\kwd{conformer}
\kwd{sound recognition}
\kwd{audio tagging}
\kwd{sound event detection}
\kwd{multimodal data}
\kwd{audiovisual data}
\kwd{missing data}
\end{keyword}


\end{abstractbox}
\end{fmbox}

\end{frontmatter}


\section{Introduction}
\label{sect:int}
Numerous sounds carry meaning relevant to everyday life: Speech is a particularly important subclass, but more general acoustic events, such as the screaming of a baby or the ringing of an alarm bell, can obviously also be of great importance to humans. In this light, it is only logical that sound recognition tasks are quickly becoming significant machine learning subjects. 

The most prominent related topics are aggregated in a yearly contest, the Detection and Classification of Acoustic Scenes and Events (DCASE) challenge. Its most recent version~\cite{DCASE2021proceedings} featured multiple subtasks revolving around classification of environmental events: If coupled with estimation of temporal boundaries, this problem is usually called sound event detection, else, it is typically referred to as audio tagging. Other subjects include but are not limited to spatial localization and automated captioning of auditory inputs. 

The research presented in this project deals with the integration of visual information into sound recognition systems. Prior efforts have shown that this process can lead to enhanced predictions: \cite{parekh2019weakly, boes2019audiovisual, yin2021enhanced} have shown that employing audiovisual variants of deep learning architectures, such as feedforward and attention-based neural networks, can be beneficial for audio tagging. In~\cite{boes2021audiovisual}, it is shown that applying early (feature) fusion to convolutional recurrent models can provide improvements for sound event detection as well. 

Many commonly used data sets for sound recognition tasks in some way stem from Audio Set~\cite{gemmeke2017audio}, which is a very large-scale cluster of auditory segments originated from YouTube. These clips are weakly labeled in the sense that the semantic categories of occurring audio events are given, but other details, such as their on- and offsets, are excluded. Derivations usually retain only a manually controlled subset of the full collection and occasionally append extra information: As an example, AudioCaps~\cite{kim2019audiocaps} attaches multiple descriptive text captions to each comprised sample.

As a consequence of much sound recognition data having its origin in YouTube, a multimedia library, it is straightforward to retrieve videos and eventually incorporate these into systems designed for related tasks. This is the approach taken by works mentioned earlier in this section. However, this procedure runs into one specific issue: Content may have been removed from this platform at some point in time, e.g., because of account deletions or copyright claims. Consequentially, the source material can not necessarily be utilized to perform computations for all samples of the considered collection. On the acoustic side, this is usually not a problem: Curators of data sets derived from Audio Set~\cite{gemmeke2017audio} often ensure availability of all audio snippets by maintaining a separate database of copies. Having said that, for the visual information, this kind of careful conservation is unfortunately not customary.

This situation can be regarded as an intense expression of the problem of missing data. As outlined in~\cite{tlamelo2021survey}, one possible approach to this matter is to simply discard all data entries with absent values. This is also the route that has been taken in the previously referenced projects related to sound recognition with visual assistance. Another, less destructive option involves imputation or replacement of missing entries. This technique can be utilized in conjunction with deep learning models, such as generative adversarial networks~\cite{ramachandram2017deep}, and currently occupies most of the literary space. Applications can be found in, among others, the domains of healthcare~\cite{le2018comparison}, biomedical data mining~\cite{petrazzini2021evaluation}, recommendation systems~\cite{lee2018impute} and speech recognition~\cite{kafoori2018robust}.

The downside to imputation techniques is that, in order for them to work properly, they require quite strong assumptions about the statistics of the complete data~\cite{little2021missing}. For example, many algorithms are based on the so-called missing at random condition: In this hypothesis, the absent values are connected to the known entries through variables which may or may not be hidden. This conjecture can undoubtedly be justified in some cases, such as well-structured time series, but is definitely not universally applicable.

In the context of this work, these assumptions are hard to defend: We deal with noisy audiovisual clips for which the two modalities are relatively decoupled. There are two frequently occurring issues in this regard. Firstly, while the sound is guaranteed to be accurately labeled, the curators of the used data set make no such promises for the visual component. For instance, the videos of some samples consist of nothing more than a meaningless still image or even a black screen. Secondly, even for the examples with real visual information, there is often a severe lack of semantic or temporal synchronization between the two streams.

Nevertheless, in this project, we still seek to tackle this specific manifestation of the problem of missing values related to sound recognition based on audiovisual clips originated from YouTube. To this end, we stray away from the unsatisfactory deletion procedure, which has been employed in previous works on the considered subject, and imputation, which is unrealistic in this instance as explained before. Instead, we propose an approach based on dynamically weighted fusion of intermediary auditory and visual features. This is done by adapting the multi-encoder framework presented in~\cite{lohrenz2021multiencoder}, which was originally utilized to achieve more robust predictions for speech recognition, without necessarily increasing time and/or memory complexity.

This work is organized as follows: In Section~\ref{sect:method}, we elaborate upon the proposed method. In Section~\ref{sect:exp}, we go into the performed experiments: We provide a detailed description of the used setup and analyze obtained results. Finally, in Section~\ref{sect:con}, we summarize the most important conclusions of the conducted research.

\section{Method}
\label{sect:method}
In this section, the proposed method is discussed. We adapt and extend the middle fusion approach presented in~\cite{lohrenz2021multiencoder} to produce a process that allows us to combine auditory inputs with partially available visual features, during training as well as inference. Section~\ref{sect:att} goes into the attention-based neural networks which are employed as base components of the considered systems. Next, in Section~\ref{sect:multienc}, the multimodal multi-encoder learning framework is fully explained.

\subsection{Attention-based architectures}
\label{sect:att}

In this section, we elaborate upon the attention-based neural networks which are used throughout this work. First, we provide a detailed description of the transformer~\cite{vaswani2017attention}, which at first was used to tackle machine translation but has since been used for many other purposes. Afterwards, we examine the conformer model~\cite{gulati2020conformer}, which augments the previously mentioned architecture with convolutional blocks and was originally designed for automatic speech recognition. In the outline below, these systems are presented in concrete forms suitable for the tasks relevant to this work, i.e., audio tagging and sound event detection. 

\subsubsection{Transformer model}
\label{subsubsect:transformer}

\begin{figure*}[!ht]
  \includegraphics{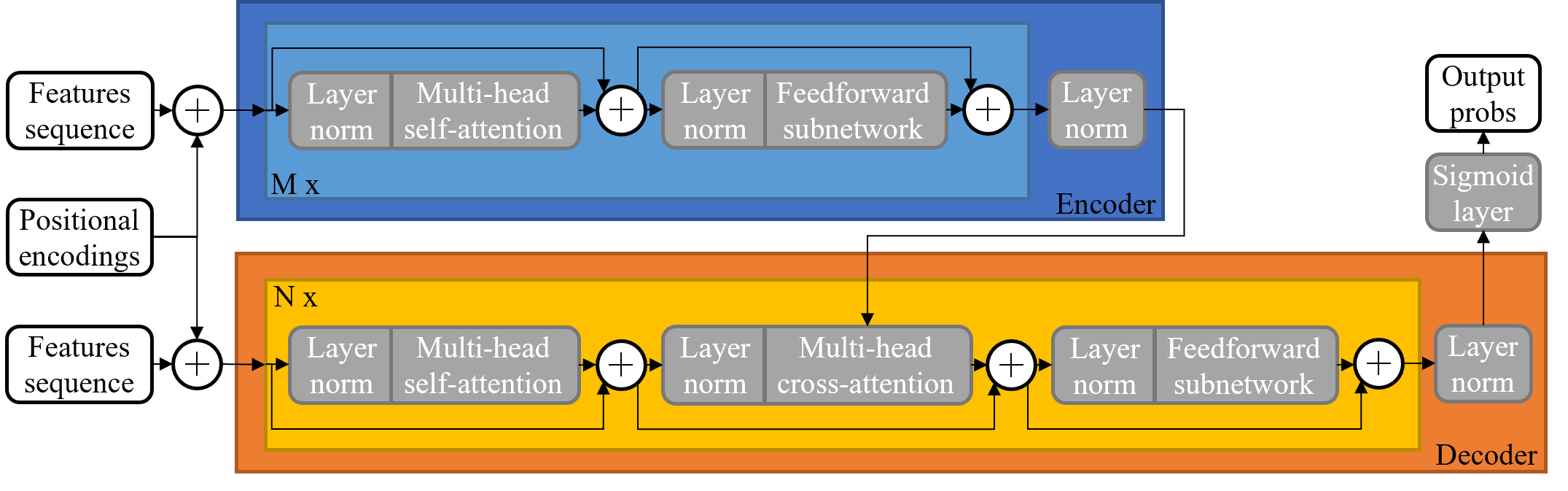}
  \caption{Diagram of transformer model for sound recognition}
  \label{fig:transformer}
\end{figure*}

The transformer for joint audio tagging and sound event detection is schematically illustrated in Figure~\ref{fig:transformer}. It strongly resembles the system described in~\cite{xiong2020layer}, which itself is a variant of the original model~\cite{vaswani2017attention}. 

The transformer is a neural network that converts sequential inputs into a series of probabilities. In the context of this project, these values indicate which sounds are present during each time frame of an audio(visual) recording. Similar to the approach taken in the BERT model~\cite{kenton2019bert}, built for natural language processing tasks, we append a learnable classification token to the set of features at the decoder side. Because of this change, the output of the system will contain an extra vector, which can be used to obtain clip-level predictions.

The encoder and decoder blocks of the transformer model (in this project, we use 3 of each) closely resemble each other. Their architectures consist of a combination of layer normalization operations~\cite{ba2016layer}, residual connections~\cite{he2016deep}, feedforward components --- which, in this case, are made up of two consecutive layers with 512 ReLU and 128 linear neurons respectively --- and perhaps most importantly, multi-head attention modules. Also, dropout~\cite{srivastava2014dropout} with a rate of 0.1 is used after each of the aforementioned feedforward submaps, this is not explicitly shown in Figure~\ref{fig:transformer}.

The multi-head attention mechanism~\cite{vaswani2017attention} performs a content-based comparison between two sets of features, referred to as queries and keys respectively. This is done by computing scaled dot products. Afterwards, the resulting so-called attention weights are multiplied with another series of inputs, namely, the values. This operation is mathematically expressed in Equation~\eqref{eq:att}.

\begin{equation}
A(Q, K, V) = \text{softmax}\left(\dfrac{QK^T}{\sqrt{d_k}}\right)V
\label{eq:att}
\end{equation}

In Equation~\eqref{eq:att}, $Q$, $K$ and $V$ are the matrices containing queries, keys and values respectively. The scaling parameter in the denominator of the softmax function, $d_k$, refers to the size of the keys. In this project, we ensure this number is equal to 128 at all times.

The multi-head module extends the simple attention calculation given in Equation~\eqref{eq:att} by applying linear mappings to the queries, keys and values before this computation and repeating the entire process several times. Eventually, the outcomes of the so-called multiple (in this work, we stick to 3) attention heads are concatenated and fed through a last projection layer with 128 units to obtain the final result. A practical detail to be added to this description is that dropout~\cite{srivastava2014dropout} with a rate of 0.1 is also applied to all attention weights and outputs of the considered neural network blocks.

All components of the transformer model that have been discussed this far perform content-based computations and ignore sequential information. This shortcoming is mitigated by adding learnable positional encodings~\cite{gehring2017convolutional} to the inputs: These are trainable vectors representing the absolute (temporal) locations of all frames in the used feature sequences.

Previously, transformer encoders have successfully been used to tackle sound event detection~\cite{miyazaki2020conformer}. When only one set of inputs is employed, as is the case for the cited work, it does indeed not make sense to utilize the full structure: The addition of a decoder would not add functionality but only increase model complexity. However, in this project, we attempt to exploit multiple sequences of features, and in that case, using the complete transformer as a basis is more appropriate. This is further elaborated upon in Section~\ref{sect:multienc}.

\subsubsection{Conformer model}

\begin{figure*}[!ht]
  \includegraphics{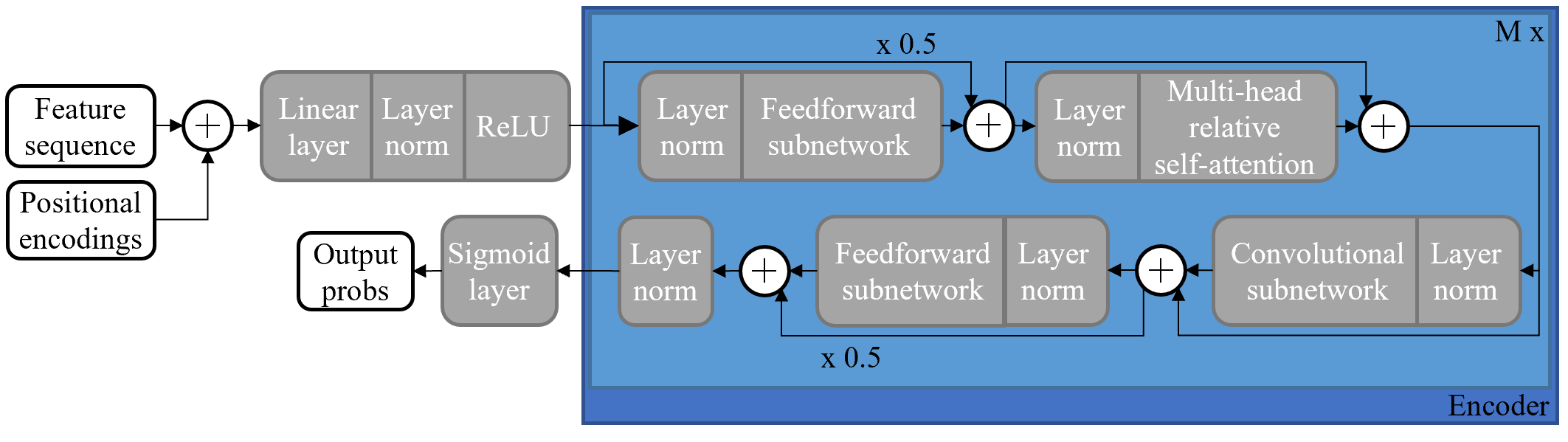}
  \caption{Diagram of conformer encoder model for sound recognition}
  \label{fig:conformerenc}
\end{figure*}

The conformer encoder~\cite{gulati2020conformer} is an extension of the corresponding component in the transformer. Compared to the latter architecture, which focuses on finding global dependencies between data frames, modules are added to substantially improve its ability to perform local processing. Figure~\ref{fig:conformerenc} depicts an appropriately designed model~\cite{miyazaki2020conformer} based on this deep learning entity for joint audio tagging and sound event detection.

A full explanation of all components in this model can be found in the original work~\cite{gulati2020conformer}. In what follows, we mainly focus on the surface-level differences between transformer and conformer encoders, while ignoring some less important and finer details. 

Firstly, the feedforward submodule in the transformer encoder is replaced by two such structures in the conformer variant. The associated residual connections are foreseen of a halving operation, akin to the Macaron neural network proposed in~{\cite{lu2020understanding}}. Also, instead of a rectified linear unit (ReLU), the more complex Swish activation function~\cite{ramachandran2017searching} is employed.

Secondly, the multi-head attention block explained in the previous section is exchanged for a version that also incorporates relative positional encodings~\cite{dai2019transformer}: By injecting these (learnable) embeddings, representing relative distances between feature vectors, the ability of this module to perform location-based (as opposed to content-based) processing is notably augmented.

\begin{figure}[!ht]
  \includegraphics{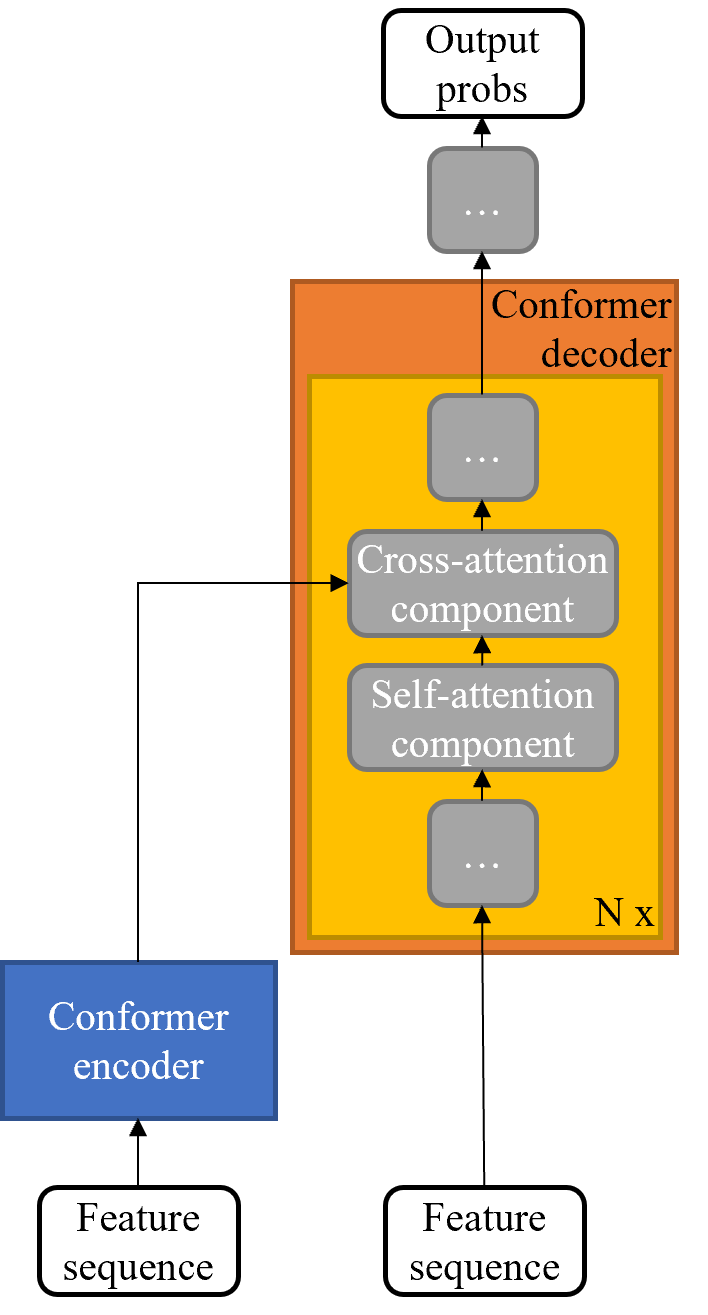}
  \caption{Simplified diagram of encoder-decoder conformer model for sound recognition}
  \label{fig:conformer}
\end{figure}

Thirdly, and perhaps most importantly, a convolutional subnetwork is inserted into the architecture. This component is what allows the conformer encoder to perform local computations, in contrast to the transformer variant, which is only capable of dealing with global (and mostly content-based) dependencies. This extra module consists of the following sequence of operations: pointwise convolution coupled with a gated linear unit~\cite{dauphin2017language}, depthwise convolution (with a kernel size of 7, as in~\cite{miyazaki2020conformer}), batch normalization~\cite{ioffe2015batch} with a momentum of 0.9, application of the Swish activation function~\cite{ramachandran2017searching}, another pointwise convolution and finally, dropout~\cite{srivastava2014dropout} with a rate of 0.1.

Conformer encoders were originally designed for automatic speech recognition but, as demonstrated, have also been employed for sound recognition~\cite{miyazaki2020conformer}. This neural structure cannot deal with more than one series of inputs, which is necessary for our purposes. Luckily, it can easily be expanded into a more appropriate encoder-decoder architecture by following the blueprint of the transformer model shown in Figure~\ref{fig:transformer}.

In the rest of this work, when we mention the conformer system, we refer to the model of which a simplified diagram is drawn in Figure~{\ref{fig:conformer}}. It is structurally similar to the architecture in Figure~{\ref{fig:transformer}}, but instead of transformer-based components, it utilizes a conformer encoder and decoder. The latter building block is derived from the former by simply adding a multi-head relative cross-attention module (with corresponding layer normalization and residual connection) after the self-attention variant, analogously to the transformer.

\subsection{Multimodal multi-encoder learning framework}
\label{sect:multienc}

In this section, we discuss the proposed multimodal multi-encoder learning framework. To this end, we first review the original middle fusion algorithm described in~\cite{lohrenz2021multiencoder}. Afterwards, we explain the adjustments that have to be made to ensure the resulting approach is suited to tackle the problem at hand, i.e., sound recognition with partial visual assistance.

\begin{figure}[!ht]
  \includegraphics{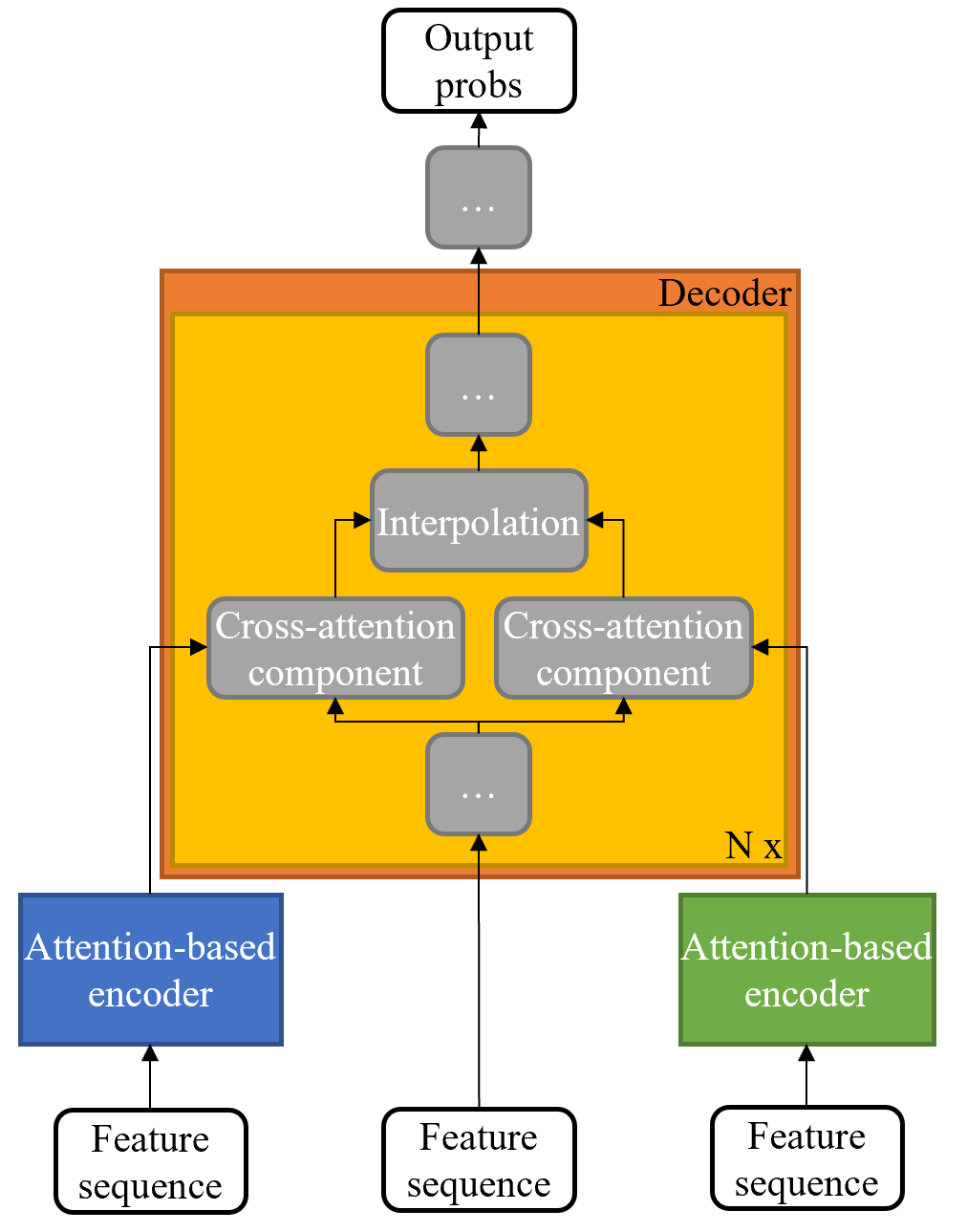}
  \caption{Simplified diagram of multi-encoder framework}
  \label{fig:multienc}
\end{figure}

Figure~\ref{fig:multienc} shows a simplified diagram of the middle fusion multi-encoder learning framework. Compared to the base architectures discussed in Section~\ref{sect:att}, the cross-attention modules of all decoder blocks are duplicated a number of times, depending on the amount of input sequences supplied to corresponding encoder structures. The vectors produced by these copies are interpolated in a linear fashion to obtain intermediate features which are used downstream in the model.

In~\cite{lohrenz2021multiencoder}, this principle is applied in the context of automatic speech recognition: Features representing the spectral magnitude and phase components of the used auditory data are combined using this multi-encoder approach to enhance the performance of the decoder, which is charged with the task of transforming input characters into output token probabilities. Fixed interpolation weights are used for calculating the relevant convex sums, biased towards the (generally) more salient magnitude representations. The cited work also investigates other configurations, such as a late fusion version and a variant with tied encoder parameters, which can be used to limit memory complexity. Preliminary experiments have shown that these options do not provide additional benefits in the context of this project, and thus, they are not discussed further.

As explained in Section~{\ref{sect:int}}, we want to build a system that can perform sound recognition using multimodal data, of which the auditory component is always at hand, but the visual information is only partially available. To this end, we adopt the framework depicted in Figure~\ref{fig:multienc}, but in contrast to the original approach, we set the multi-encoder interpolation parameters dynamically, both during training and testing: When series of features, i.e., those related to vision, are missing, the weights associated with these sequences are set to 0. Naturally, the others are scaled in order for the total to remain equal to 1. In Section~\ref{sect:exp}, we also explore a novel weighting scheme for the learning phase.

\section{Experiments}
\label{sect:exp}

We detail the performed experiments involving the proposed multi-encoder framework, elaborated upon in Section~\ref{sect:method}, meant to deal with the considered problem of missing values. In Section~\ref{subsect:setup}, the setup is discussed, while Section~\ref{subsect:results} analyzes the obtained results.

\subsection{Setup}
\label{subsect:setup}

In this section, we fully lay out the experimental setup. Firstly, the used data set and the features that are extracted from this collection are elaborated upon. Next, we report the postprocessing steps applied to convert the output probabilities of the models described in Section~\ref{sect:method} into binary predictions, and we go into the metrics employed to gauge the performance of the examined systems. Finally, for completeness, we list all relevant training and testing hyperparameters.

\subsubsection{Data}

In this work, we use the data associated with task 4~{\cite{turpault2019sound}} of the DCASE 2020 challenge~{\cite{DCASE2020proceedings}}. This large-scale set contains recordings with a maximum length of 10 seconds. Each instance features a number of potentially overlapping auditory events belonging to the 10 possible environmental classes itemized in Table~{\ref{tab:class}}. The collection is split into multiple partitions. The amount of samples in each subset is listed in Table~{\ref{tab:data}}.

\begin{table}[hbt!]
\caption{Labeled sound categories in data set}
\label{tab:class}
\begin{tabular}{@{}ll@{}}
\toprule
Speech & Frying \\
Dog & Blender \\
Cat & Running water \\
Alarm/bell/ringing & Vacuum cleaner \\
Dishes & Electric shaver/toothbrush \\
\bottomrule
\end{tabular}
\end{table}

Details about the occurring auditory events are provided in different ways for the distinct collections mentioned in Table~{\ref{tab:data}}. The strongly supervised training, validation and evaluation partitions encompass full annotations of all occurring sounds, including on- and offsets. The samples of the other subsets are either weakly supervised, i.e., only clip-level labels are available, or simply do not contain any relevant information at all.

\begin{table}[hbt!]
\caption{Number of samples per partition in data set}
\label{tab:data}
\begin{tabular}{@{}lc@{}}
\toprule
\textbf{Data partition} & \textbf{Number of recordings} \\
\midrule
Unsupervised training & 14412 \\
Weakly supervised training & 1578 \\
Strongly supervised training & 2584 \\
Validation & 1168 \\
Evaluation (public) & 692 \\
\bottomrule
\end{tabular}
\end{table}

The recordings in the strongly supervised training set originate from Freesound, a collaborative database of sounds, and only contain auditory information. All other clips come from YouTube, a very well-known large-scale multimedia library. For most of these examples, visual information can also be extracted. However, as thoroughly explained in Section~\ref{sect:int}, this is not possible for all data points: On average, for 15.5\% of those samples, the video can no longer be downloaded.

\subsubsection{Features}

In this project, we investigate multiple ways of preprocessing the auditory and visual streams of the considered data. In Section~\ref{subsect:results}, the discussion on the results of the experiments includes an analysis of which features lead to improvements under the proposed framework for a number of evaluation scenarios. 

\paragraph{Spectral auditory features}

We first resample the audio streams to 16 kHz and apply peak amplitude normalization. Next, log mel magnitude spectrograms with 64 frequency bins are extracted using a window size of 1024 samples (corresponding to 64 ms) and a hop length of 313 samples (corresponding to 20 ms). For a clip of 10 seconds, this results in 512 frames. Lastly, per frequency bin standardization is performed, based on the statistics of the training data.

Compared to the other (pretrained) embeddings described below, these features are fairly rudimentary. To create higher-level auditory vectors which can be supplied as inputs to the models described in Section~\ref{sect:method}, we use a feature extractor with a similar architecture as in the baseline model~\cite{turpault2020training} for task 4~\cite{turpault2019sound} of the DCASE 2020 challenge~\cite{DCASE2020proceedings}. It is a convolutional network which is made up of seven consecutive stacks, which each perform the following five operations: convolution, batch normalization~\cite{ioffe2015batch} with a momentum of 0.9, application of the ReLU function, dropout~\cite{srivastava2014dropout} with a rate of 0.1 and average pooling. The hyperparameters of the convolutional layers are listed per block in Table~\ref{tab:conv}.

\begin{table}[hbt!]
\caption{Hyperparameters of convolutional feature extractor}
\label{tab:conv}
\begin{tabular}{@{}llccc@{}}
\toprule
\textbf{Block} & \textbf{Operation} & \textbf{Channels} & \textbf{Kernel size} & \textbf{Strides} \\
\midrule
\multirow{2}{*}{0} & Convolution  & 16  & (3, 3)  & (1, 1) \\
 & Pooling & - & (2, 2)  & (2, 2) \\
\midrule
\multirow{2}{*}{1} & Convolution  & 32  & (3, 3)  & (1, 1) \\
 & Pooling & - & (2, 2)  & (2, 2) \\
\midrule
\multirow{2}{*}{2} & Convolution  & 64  & (3, 3)  & (1, 1) \\
 & Pooling & - & (2, 2)  & (2, 2) \\
\midrule
\multirow{2}{*}{3-4-5-6} & Convolution  & 128  & (3, 3)  & (1, 1) \\
 & Pooling & - & (1, 2)  & (1, 2) \\
\bottomrule
\end{tabular}
\end{table}

In Table~\ref{tab:conv}, the first and second numbers of each tuple in the columns on kernel size and stride relate to time and frequency axes respectively. As can be inferred from this list, this convolutional feature extractor reduces the frequency dimension of the spectral map to one and has a total temporal pooling factor of 8: For a clip of 10 seconds, this results in the original series being transformed into a sequence of 64 vectors.

This convolutional feature extractor is inserted into the models described in Section~\ref{sect:method} and the complete architecture is trained in an end-to-end manner.

\paragraph{OpenL3 visual features}

OpenL3~\cite{cramer2019look} is an embedding model designed to predict correspondence between auditory and visual streams, trained in a self-supervised way. It is pretrained on Audio Set~\cite{gemmeke2017audio}. 

To obtain pretrained visual features, still images are first sampled from the available visual streams at a rate of about 6.5 fps. The frames are fed into the video subnetwork of OpenL3~\cite{cramer2019look}. For a recording of 10 seconds, these steps lead to 64 512-dimensional vectors. 

\paragraph{Temporally coherent visual features}

Still images are first sampled from the visual streams at a rate of about 6.5 fps. Next, these frames are passed through the video embedding model described in~{\cite{knights2021temporally}}, trained in a self-supervised manner using a loss optimizing temporal coherency. For a clip of 10 seconds, this procedure results in a series of 64 2048-dimensional vectors. 

\paragraph{VGG16 visual features}

Still images are sampled from the visual streams at a rate of about 6.5 fps. The resulting frames are fed into VGG16~\cite{simonyan2015very}, a convolutional model for image classification, pretrained on the ImageNet data set~\cite{russakovsky2015imagenet}. The 4096-dimensional outputs of the last feedforward layer of this neural network are used as feature sequences in this project. For a recording of 10 seconds, these steps lead to 64 vectors. 

\subsubsection{Postprocessing}

To calculate clip-level metrics, the relevant probabilities are converted into binary decisions by employing class-wise thresholds, optimized on the validation partition of the employed data set. The search space for these hyperparameters is restricted to the linear span ranging from 0.1 to 0.9 in steps of 0.1.

To obtain sound event detection scores, the frame-level probabilities are transformed via the following process: Firstly, they are binarized, and secondly, the decisions are passed through a median smoothing operation. The hyperparameters associated with these steps are (separately) optimized on the validation partition of the employed data set, per sound category as well as per used evaluation metric. The search space for the thresholds is limited to a linear span ranging from 0.1 to 0.9 in steps of 0.1. For the filter sizes, values from 1 to 31 are tested in increments of 2.

Other postprocessing operations can be applied to further enhance the predictions of the considered systems. For example, in~\cite{miyazaki2020conformer}, model ensembling is employed in the form of score-level fusion. In this work, we choose not to take any further steps as the number of possibilities is seemingly unlimited. Also, the main goal of this project is not necessarily to achieve the perfectly optimized model, rather, we want to demonstrate the effectiveness of the proposed method.

\subsubsection{Metrics}

To evaluate the considered models for both audio tagging and sound event detection, we use a variety of metrics representing distinct scenarios, involving different requirements in terms of temporal localization. These scores are specifically chosen because of their prevalence in the field of sound recognition.

\paragraph{Clip-based F1 score (CBF1)}

For audio tagging, we use the micro-averaged clip-based F1 score~\cite{mesaros2016metrics}. This metric was also used in the DCASE 2017 challenge~\cite{DCASE2017proceedings}.

\paragraph{Segment-based F1 score (SBF1)}

We use the micro-averaged segment-based F1 score based on slices of 1 s~\cite{mesaros2016metrics} to quantify the effectiveness of the predictions of the proposed models. Because of the relatively long slice length, this metric is suitable for investigating coarse-grained sound event detection performance. It was also utilized in the DCASE 2017 challenge~\cite{DCASE2017proceedings}.

\paragraph{Event-based F1 score (EBF1)}

We use the macro-averaged event-based F1 score with tolerances of 200 ms for onsets and 20\% of the lengths of the audio events (up to a maximum of 200 ms) for offsets~\cite{mesaros2016metrics}. Because of the strict localization requirements, this metric is suitable for investigating fine-grained sound event detection performance. It was also utilized in the DCASE 2018~\cite{DCASE2018proceedings}, 2019~\cite{DCASE2019proceedings} and 2020~\cite{DCASE2020proceedings} challenges.

\paragraph{Polyphonic sound event detection scores (PSDS)}

\begin{table}[hbt!]
\caption{PSDS hyperparameters}
\label{tab:psds}
\centering
\begin{tabular}{@{}lcc@{}}
\toprule
\textbf{Hyperparameter} & \textbf{PSDS1} & \textbf{PSDS2} \\
\midrule 
Detection tolerance criterion & 0.7 & 0.1 \\
Ground truth intersection criterion & 0.7 & 0.1 \\
Cross-trigger tolerance criterion & N/A & 0.3 \\
Cost of class instability & 1 & 1 \\
Cost of cross-triggers & 0 & 0.5 \\
Maximum false positive rate & 100 & 100 \\ 
\bottomrule
\end{tabular}
\end{table}

We use two polyphonic sound event detection scores~\cite{bilen2020framework}, representing distinct evaluation scenarios. In the rest of this work, they are referred to as PSDS1 and PSDS2. The former imposes strict requirements on the temporal localization accuracy, the latter is more lenient in this regard. The hyperparameters for these measures are summarized in Table~\ref{tab:psds}. One of the default postprocessing steps described before is left out in this case: These scores are computed using 50 fixed operating points, in which thresholds linearly distributed from 0.01 to 0.99 (with a step size of 0.02) are used to convert probabilities into binary decisions. These metrics were also utilized in the DCASE 2021~\cite{DCASE2021proceedings} challenge.

\subsubsection{Hyperparameters}

In this section, we provide details on the used training and testing procedures for the sake of completeness. PyTorch~\cite{paszke2019pytorch} is utilized to implement all of the work. 

Elaborate explanations of the selection processes for the encoder interpolation weights of the considered models are omitted from this section, as these procedures vary per experimental setting. Instead, these descriptions are left for the relevant parts in Section~{\ref{subsect:results}}.

\paragraph{Training}

As already outlined in Section~{\ref{subsect:setup}}, the data employed in this research project is heterogeneously annotated, and thus, combining all available samples into the learning procedure is challenging, especially when it comes to the unlabeled instances. To deal with this difficulty, mean teacher training~{\cite{tarvainen2017mean}} is performed. In this framework, two models called the student and the teacher are utilized. They share the same architecture, but their parameters are updated differently.

The student system is trained regularly, i.e., a differentiable objective is minimized. However, the weights of the teacher are computed as the exponential moving average of the student parameters with a multiplicative decay factor of 0.999 per training iteration.

The loss employed to train the student consists of four terms: The first two are clip-level and frame-level binary cross entropy functions, which are only computed for the weakly and strongly labeled clips respectively. The other components are mean-squared error consistency costs between the clip-level and frame-level output probabilities of the student and teacher models, which can be computed for all data, including the unannotated samples. The classification and consistency terms are summed with weights 1 and 2 respectively to obtain the final objective.

\begin{table*}[hbt!]
\caption{Results of uni- and bi-encoder attention-based models}
\label{tab:unibi}
\centering
\begin{tabular}{@{}llllccccc@{}}
\toprule
\multirowcell{2}[-0.4ex][l]{\textbf{Base model}} & \multirowcell{2}[-0.4ex][l]{\textbf{Encoder inputs}} & \multicolumn{2}{l}{\textbf{Encoder weights}} & \multirowcell{2}[-0.4ex]{\textbf{CBF1} \\ \textbf{(\%)}} & \multirowcell{2}[-0.4ex]{\textbf{SBF1} \\ \textbf{(\%)}} & \multirowcell{2}[-0.4ex]{\textbf{EBF1} \\ \textbf{(\%)}} & \multirowcell{2}[-0.4ex][l]{\textbf{PSDS1}} & \multirowcell{2}[-0.4ex][l]{\textbf{PSDS2}} \\
\cmidrule(l){3-4}
& & \textbf{Training} & \textbf{Inference} & & &  & & \\
\midrule 
\multirowcell{8}[-0.8ex][l]{Transformer} & \multirowcell{2}[0ex][l]{Spectral auditory features} & \multirowcell{2}[0ex][l]{1} & \multirowcell{2}[0ex][l]{1} & 82.48 & 78.92 & 49.43 & 0.4083 & 0.6413 \\
 & & & & \textpm 0.47 & \textpm 0.48 & \textpm 0.61 & \textpm 0.0080 & \textpm 0.013 \\
\cmidrule(l){2-9}
 & Spectral auditory features & 0.5 & 0.75 & \textbf{84.57} & \textbf{80.56} & \textbf{51.00} & 0.4168 & \textbf{0.6567} \\
 & OpenL3 visual features & 0.5 & 0.25 & \textpm 0.40 & \textpm 0.40 & \textpm 0.55 & \textpm 0.0080 & \textpm 0.011 \\
\cmidrule(l){2-9}
& Spectral auditory features & 0.75 & 0.875 & 84.10 & 80.11 & 50.97 & \textbf{0.4188} & 0.6502 \\
 & Temporally coherent visual features & 0.25 & 0.125 & \textpm 0.45 & \textpm 0.46 & \textpm 0.58 & \textpm 0.0086 & \textpm 0.014 \\
\cmidrule(l){2-9}
& Spectral auditory features & 0.5 & 0.875 & 84.22 & 79.91 & 48.28 & 0.3891 & 0.6144 \\
 & VGG16 features & 0.5 & 0.125 & \textpm 0.41 & \textpm 0.41 & \textpm 0.60 & \textpm 0.0079 & \textpm 0.011 \\
\midrule 
\multirowcell{8}[-0.8ex][l]{Conformer} & \multirowcell{2}[0ex][l]{Spectral auditory features} & \multirowcell{2}[0ex][l]{1} & \multirowcell{2}[0ex][l]{1} & 82.52 & 78.62 & 48.41 & 0.3849 & 0.6451 \\
 & & & & \textpm 0.40 & \textpm 0.41 & \textpm 0.57 & \textpm 0.0086 & \textpm 0.011 \\
\cmidrule(l){2-9}
 & Spectral auditory features & 0.75 & 0.875 & 83.83 & 80.22 & \textbf{50.41} & \textbf{0.4010} & \textbf{0.6501} \\
 & OpenL3 visual features & 0.25 & 0.125 & \textpm 0.44 & \textpm 0.45 & \textpm 0.64 & \textpm 0.0090 & \textpm 0.012 \\
\cmidrule(l){2-9}
& Spectral auditory features & 0.75 & 0.875 & \textbf{84.15} & \textbf{80.48} & 50.18 & 0.4000 & 0.6490 \\
 & Temporally coherent visual features & 0.25 & 0.125 & \textpm 0.42 & \textpm 0.38 & \textpm 0.63 & \textpm 0.0092 & \textpm 0.011 \\
\cmidrule(l){2-9}
& Spectral auditory features & 0.5 & 0.875 & 83.83 & 79.43 & 46.48 & 0.3615 & 0.6226 \\
 & VGG16 features & 0.5 & 0.125 & \textpm 0.43 & \textpm 0.48 & \textpm 0.60 & \textpm 0.0079 & \textpm 0.014 \\
\bottomrule
\end{tabular}
\end{table*}

During training, data augmentation is also employed in the form of mixup~\cite{zhang2018mixup}, which comes down to creating extra learning examples (and associated labels) by linearly interpolating the original samples. We use this method with an application rate of 33\%. The mixing ratios are randomly sampled from a beta distribution with shape parameters equal to 0.2.

Models are trained for 100 epochs. Per epoch, 250 batches of 128 samples are given to the networks. Each batch contains 32 strongly labeled, 32 weakly labeled and 64 unlabeled examples. Rectified Adam~\cite{kingma2015adam, liu2019variance} is employed to train the weights of the student systems. Learning rates start at 0.001 and decay multiplicatively with a factor of 0.1 per 10000 iterations.

\paragraph{Testing}

Metrics are calculated on probabilities produced by student models after the last training epoch.

\subsection{Results}
\label{subsect:results}

In this section, all experimental results are listed and analyzed. We report evaluation scores which have been averaged over 20 training runs (with independent initializations of all model parameters) to ensure reliability, as well as the associated standard deviations.

Preliminary experiments have indicated that architectures only employing visual features underperform badly with regard to sound event detection. These outcomes are in line with findings divulged in prior research~\cite{boes2021audiovisual}. Clearly, on their own, these pretrained vectors are not able to properly perform temporal segmentation on the considered multimodal data. As a consequence of this observation, the following design choice has been made: All of the explored systems take spectral auditory maps as inputs to their decoders and to one of their encoders. In other words, in what follows, models without acoustic features are not considered.

\subsubsection{Uni-encoder attention-based models}

Table~\ref{tab:unibi} contains the results obtained by baseline models which do not utilize any visual information at all and thus do not run into the examined missing data problem: Each of these systems uses auditory features as inputs to its decoder as well as its single encoder.

In~\cite{miyazaki2020conformer}, transformer and conformer encoders have been used in a very similar way to tackle sound event detection on the same data set. The performance values reported in the cited work for such models that do not use any type of ensemble method are comparable to those in Table~\ref{tab:unibi}. The remaining small disparities can partially be attributed to architectural differences, since the referenced systems do not include decoder components, unlike is the case in this project.

Interestingly, models using the transformer architecture as a base outperform those using conformer components in terms of most considered performance metrics, in contrast to what is reported in~\cite{miyazaki2020conformer}. This trend will reappear in the results of systems exploiting visual information (or equivalently, using multiple encoders), which are discussed in the following sections. However, the differences are too small and/or inconsistent to infer conclusions on the superiority of one or the other.

\subsubsection{Bi-encoder attention-based models}

Table~\ref{tab:unibi} also lists the scores obtained by models that use two encoders: one takes in spectral auditory maps, the other accepts a type of pretrained visual features. 

The interpolation weights associated with these two encoders are determined by maximizing the performance of the models on the validation partition of the data set at hand. Specifically, we choose the hyperparameters which lead to the highest event-based F1 scores, and empirically, we find that this also leads to near-optimal values in terms of the other considered evaluation metrics. For the encoder weight associated with acoustic information, we test the following options during training as well as inference: 0.25, 0.5, 0.75 and 0.875. When visuals are unavailable, this number is set to 1, as explained in Section~\ref{sect:multienc}.

\paragraph{Discussion on pretrained visual features}
A comparison between the results of the bi-encoder systems incorporating visual information and the scores obtained by the baseline models only utilizing acoustic inputs, both listed in Table~{\ref{tab:unibi}}, demonstrates that the proposed method can be useful, but the choice of pretrained vision-related vectors is crucial. Particularly, we observe that adding VGG16 embeddings does not globally lead to improvements, but only for clip- and segment-based F1 scores. However, the inclusion of OpenL3 and temporally coherent features does provide consistent and substantial performance boosts.

\begin{table*}[hbt!]
\caption{Results of bi-encoder attention-based models with random encoder interpolation weights during training}
\label{tab:birandom}
\centering
\begin{tabular}{@{}lllccccc@{}}
\toprule
\multirowcell{2}[-0.4ex][l]{\textbf{Base model}} & \multirowcell{2}[-0.4ex][l]{\textbf{Encoder inputs}} & \textbf{Encoder weight} & \multirowcell{2}[-0.4ex]{\textbf{CBF1} \\ \textbf{(\%)}} & \multirowcell{2}[-0.4ex]{\textbf{SBF1} \\ \textbf{(\%)}} & \multirowcell{2}[-0.4ex]{\textbf{EBF1} \\ \textbf{(\%)}} & \multirowcell{2}[-0.4ex][l]{\textbf{PSDS1}} & \multirowcell{2}[-0.4ex][l]{\textbf{PSDS2}} \\
\cmidrule(l){3-3}
& & \textbf{Inference} & & & & & \\
\midrule 
\multirowcell{6}[-0.8ex][l]{Transformer} & Spectral auditory features & 0.75 & 84.20 & \textbf{80.71} & \textbf{50.80} & 0.4169 & \textbf{0.6557} \\
 & OpenL3 visual features & 0.25 & \textpm 0.39 & \textpm 0.48 & \textpm 0.65 & \textpm 0.0089 & \textpm 0.010 \\
\cmidrule(l){2-8}
& Spectral auditory features & 0.875 & 84.42 & 80.46 & 50.64 & \textbf{0.4201} & 0.6554 \\
 & Temporally coherent visual features & 0.125 & \textpm 0.37 & \textpm 0.39 & \textpm 0.58 & \textpm 0.0079 & \textpm 0.013 \\
\cmidrule(l){2-8}
& Spectral auditory features & 0.875 & \textbf{84.48} & 80.20 & 46.95 & 0.3818 & 0.6170 \\
 & VGG16 features & 0.125 & \textpm 0.47 & \textpm 0.39 & \textpm 0.55 & \textpm 0.0075 & \textpm 0.014 \\
\midrule 
\multirowcell{6}[-0.8ex][l]{Conformer} & Spectral auditory features & 0.875 & 83.94 & 80.27 & \textbf{50.00} & \textbf{0.4051} & 0.6539 \\
 & OpenL3 visual features & 0.125 & \textpm 0.39 & \textpm 0.44 & \textpm 0.64 & \textpm 0.0074 & \textpm 0.011 \\
\cmidrule(l){2-8}
& Spectral auditory features & 0.875 & 84.18 & \textbf{80.50} & 49.94 & 0.4040 & \textbf{0.6552} \\
 & Temporally coherent visual features & 0.125 & \textpm 0.36 & \textpm 0.38 & \textpm 0.57 & \textpm 0.0078 & \textpm 0.013 \\
\cmidrule(l){2-8}
& Spectral auditory features & 0.875 & \textbf{84.25} & 80.33 & 46.10 & 0.3591 & 0.6201 \\
 & VGG16 features & 0.125 & \textpm 0.39 & \textpm 0.50 & \textpm 0.67 & \textpm 0.0089 & \textpm 0.012 \\
\bottomrule
\end{tabular}
\end{table*}

These findings are in agreement with outcomes published in prior research. In \cite{boes2021audiovisual}, it is shown that adding VGG16 embeddings can lead to improvements for audio tagging and (partly) coarse-grained sound event detection, but when it comes to stricter segmentation, these vectors do not add any value. This can be explained by the fact that the system these embeddings are extracted from is designed for a problem without time-related aspects, i.e., image classification. This is not the case for OpenL3 and temporally coherent visual features, as the associated models are pretrained for tasks encompassing temporal facets, which explains the disparity discussed in the previous paragraph.

\paragraph{Discussion on metrics}
Independent model initialization through random seeding of learnable parameters does not cause a lot of inconsistency with regard to the clip- and segment-based F1 scores, which measure audio tagging and coarse-grained sound event detection performance respectively. This allows us to conclude with certainty that systems employing the proposed method outperform the baseline when it comes to these metrics. For the event-based F1 measure, targeting comparatively strict temporal segmentation, the standard deviations across training runs listed in Table~{\ref{tab:unibi}} are slightly bigger, but still small enough to be able to make useful inferences. However, the variability is significantly greater (in relative terms) for PSDS1 and especially PSDS2, which means more caution should be exercised when interpreting those results.

\paragraph{Discussion on interpolation weights}
We find that changing the encoder weights employed while training causes relatively limited fluctuations in terms of the considered metrics. However, modifying the inference hyperparameters can cause significant performance drops. This happens in particular when the interpolation weight for the acoustic stream is too low.

Based on this observation, we present a novel way of setting these hyperparameters. While learning, we randomize this decision process: The interpolation weight associated with the acoustic input features is sampled from a uniform distribution between 0.25 and 1 per batch. It would not be logical to let this value go all the way to 0, as in that specific case, the model would have to rely on visual information only. For reasons explained in detail in Section~\ref{sect:int}, this is not a good idea as the auditory stream is generally much more salient. For inference, the default procedure for determining these weights, involving optimization based on validation data, is retained. The results obtained when using this adapted method are presented in Table~\ref{tab:birandom}.

\begin{table*}[hbt!]
\caption{Split results of uni- and bi-encoder attention-based models}
\label{tab:unibisplit}
\centering
\begin{tabular}{@{}lllccccc@{}}
\toprule
\multirowcell{2}[0ex][l]{\textbf{Base model}} & \multirowcell{2}[0ex][l]{\textbf{Encoder inputs}} & \multirowcell{2}[0ex][l]{\textbf{Evaluation data}} & \multirowcell{2}[0ex]{\textbf{CBF1} \\ \textbf{(\%)}} & \multirowcell{2}[0ex]{\textbf{SBF1} \\ \textbf{(\%)}} & \multirowcell{2}[0ex]{\textbf{EBF1} \\ \textbf{(\%)}} & \multirowcell{2}[0ex]{\textbf{PSDS1}} & \multirowcell{2}[0ex]{\textbf{PSDS2}} \\
& & & & & & & \\
\midrule 
\multirowcell{11}[-1.2ex][l]{Transformer} & \multirowcell{2}[0ex][l]{Spectral auditory features} & Video available (but ignored) & 82.79 & 79.19 & 48.89 & 0.4026 & 0.6377 \\
 & & Video unavailable & 80.40 & 77.21 & 49.79 & 0.4358 & 0.6159 \\
\cmidrule(l){2-8}
 & \multirowcell{3}[0ex][l]{Spectral auditory features \\ OpenL3 visual features} & Video available & 85.10 & 80.94 & 50.70 & 0.4140 & 0.6583 \\
 & & Video available (but ignored) & 83.26 & 79.11 & 49.03 & 0.4073 & 0.6296 \\
  & & Video unavailable  & 81.01 & 78.16 & 51.44 & 0.4311 & 0.6154 \\
\cmidrule(l){2-8}
& \multirowcell{3}[0ex][l]{Spectral auditory features \\ Temporally coherent \\ visual features} & Video available  & 84.58 & 80.60 & 50.71 & 0.4163 & 0.6514 \\
 & & Video available (but ignored) & 82.74 & 78.98 & 49.44 & 0.4073 & 0.6248 \\
  & & Video unavailable & 80.91 & 77.03 & 51.41 & 0.4296 & 0.6099 \\
\cmidrule(l){2-8}
& \multirowcell{3}[0ex][l]{Spectral auditory features \\  VGG16 features} & Video available & 85.16 & 80.73 & 48.34 & 0.3854 & 0.6222 \\
 & & Video available (but ignored) & 79.40 & 75.78 & 45.16 & 0.3591 & 0.5897 \\
  & & Video unavailable & 77.98 & 74.67 & 48.05 & 0.3926 & 0.5526 \\
\midrule 
\multirowcell{11}[-1.2ex][l]{Conformer} & \multirowcell{2}[0ex][l]{Spectral auditory features} & Video available (but ignored) & 82.85 & 78.82 & 48.04 & 0.3811 & 0.6426 \\
 & & Video unavailable & 80.24 & 77.33 & 49.68 & 0.4210 & 0.6236 \\
\cmidrule(l){2-8}
 & \multirowcell{3}[0ex][l]{Spectral auditory features \\ OpenL3 visual features} & Video available & 84.35 & 80.70 & 50.14 & 0.3954 & 0.6477 \\
 & & Video available (but ignored) & 82.68 & 79.11 & 48.43 & 0.3878 & 0.6383 \\
 & & Video unavailable & 80.28 & 77.12 & 50.46 & 0.4256 & 0.6330 \\
\cmidrule(l){2-8}
& \multirowcell{3}[0ex][l]{Spectral auditory features \\ Temporally coherent \\ visual features} & Video available & 84.70 & 80.94 & 50.00 & 0.3970 & 0.6489 \\
 & & Video available (but ignored) & 82.38 & 78.83 & 48.28 & 0.3845 & 0.6370 \\
 & & Video unavailable & 80.45 & 77.50 & 50.87 & 0.4185 & 0.6284 \\
\cmidrule(l){2-8}
& \multirowcell{3}[0ex][l]{Spectral auditory features \\  VGG16 features} & Video available & 84.03 & 80.19 & 46.09 & 0.3605 & 0.6290 \\
 & & Video available (but ignored) & 79.02 & 74.71 & 43.91 & 0.3486 & 0.5900 \\
 & & Video unavailable & 78.35 & 74.41 & 45.55 & 0.3714 & 0.5583 \\
\bottomrule
\end{tabular}
\end{table*}

The reported scores are very comparable to those in Table~\ref{tab:unibi}, for models that do not use randomized encoder interpolation weights during training. They are not much (or in some cases, any) better, but this altered learning procedure still has a significant benefit: The hyperparameter optimization process becomes less tedious and time-consuming in this case, which is valuable from a practical point of view.

\begin{table*}[hbt!]
\caption{Results of tri-encoder attention-based models}
\label{tab:tri}
\centering
\begin{tabular}{@{}llllccccc@{}}
\toprule
\multirowcell{2}[-0.4ex][l]{\textbf{Base model}} & \multirowcell{2}[-0.4ex][l]{\textbf{Encoder inputs}} & \multicolumn{2}{l}{\textbf{Encoder weights}} & \multirowcell{2}[-0.4ex]{\textbf{CBF1} \\ \textbf{(\%)}} & \multirowcell{2}[-0.4ex]{\textbf{SBF1} \\ \textbf{(\%)}} & \multirowcell{2}[-0.4ex]{\textbf{EBF1} \\ \textbf{(\%)}} & \multirowcell{2}[-0.4ex][l]{\textbf{PSDS1}} & \multirowcell{2}[-0.4ex][l]{\textbf{PSDS2}} \\
\cmidrule(l){3-4}
& & \textbf{Training} & \textbf{Inference} & & &  & & \\
\midrule 
\multirowcell{6}[-0.4ex][l]{Transformer} & Spectral auditory features & 0.75 & 0.875 & \multirowcell{3}[0ex]{\textbf{85.00} \\ \textpm 0.37} & \multirowcell{3}[0ex]{81.10 \\ \textpm 0.45} & \multirowcell{3}[0ex]{\textbf{52.01} \\ \textpm 0.63} & \multirowcell{3}[0ex]{0.4194 \\ \textpm 0.0091} & \multirowcell{3}[0ex]{\textbf{0.6653} \\ \textpm 0.013} \\
 & OpenL3 visual features & 0.125 & 0.0625 & & & & & \\
& Temporally coherent visual features & 0.125 & 0.0625 & & & & & \\
\cmidrule(l){2-9}
& Spectral auditory features & Random & 0.75 & \multirowcell{3}[0ex]{84.86 \\ \textpm 0.36} & \multirowcell{3}[0ex]{\textbf{81.40} \\ \textpm 0.40} & \multirowcell{3}[0ex]{51.82 \\ \textpm 0.55} & \multirowcell{3}[0ex]{\textbf{0.4231} \\ \textpm 0.0086} & \multirowcell{3}[0ex]{0.6622 \\ \textpm 0.011} \\
 & OpenL3 visual features & Random & 0.125 & & & & & \\
 & Temporally coherent visual features & Random & 0.125 & & & & & \\
\midrule 
\multirowcell{6}[-0.4ex][l]{Conformer} & Spectral auditory features & 0.75 & 0.875 & \multirowcell{3}[0ex]{\textbf{84.90} \\ \textpm 0.40} & \multirowcell{3}[0ex]{\textbf{81.12} \\ \textpm 0.49} & \multirowcell{3}[0ex]{\textbf{50.91} \\ \textpm 0.59} & \multirowcell{3}[0ex]{\textbf{0.4097} \\ \textpm 0.0084} & \multirowcell{3}[0ex]{0.6612 \\ \textpm 0.012} \\
 & OpenL3 visual features & 0.125 & 0.0625 & & & & & \\
 & Temporally coherent visual features & 0.125 & 0.0625 & & & & & \\
\cmidrule(l){2-9}
& Spectral auditory features & Random & 0.875 & \multirowcell{3}[0ex]{84.57 \\ \textpm 0.41} & \multirowcell{3}[0ex]{80.99 \\ \textpm 0.41} & \multirowcell{3}[0ex]{50.90 \\ \textpm 0.61} & \multirowcell{3}[0ex]{0.4087 \\ \textpm 0.0081} & \multirowcell{3}[0ex]{\textbf{0.6647} \\ \textpm 0.012} \\
 & OpenL3 visual features & Random & 0.0625 & & & & & \\
& Temporally coherent visual features & Random & 0.0625 & & & & & \\
\bottomrule
\end{tabular}
\end{table*}

\paragraph{Discussion on merit of multi-encoder learning}

In this section, we go into the merit of the multi-encoder learning framework by further scrutinizing the performance boosts reported in Table~\ref{tab:unibi}. To this end, we first split the test data into two sets, using the availability of visual information as partitioning criterion. We report the results obtained by the uni-encoder baseline as well as the proposed (non-randomized) bimodal models on these two disjoint collections. Additionally, we disclose performance scores produced by the latter systems on the evaluation data with visual assistance when the interpolation weight for the acoustic stream is forced to 1, and vision-related inputs are ignored. These outcomes are all given in Table~\ref{tab:unibisplit}. Standard deviations are not included to maintain clarity and conserve space.

Even though the bi-encoder attention-based architectures are designed to utilize inputs from two modalities, when vision-related inputs are forcibly ignored by setting the associated linear interpolation weights to 0, these models still perform on par to the uni-encoder baseline. This trend appears when inspecting the scores obtained on both of the aforementioned evaluation data splits. The only exception to this rule are the systems with pretrained VGG16 embeddings, but this is not entirely unexpected behavior, since these vectors are inappropriate for the tasks at hand as discussed before. This finding seems to indicate that, on the condition that the employed sequences of features are selected appropriately, the multi-encoder framework can be applied relatively safely, without fear of decreasing performance, regardless of the actual degree of accessibility of visual recordings.

When looking at the performance differences between the unimodal baseline and the bi-encoder models on the evaluation data that includes visual inputs, we find that there are consistently substantial improvements --- once again, with the sole exception being systems utilizing pretrained VGG16 features. This is obviously desirable as the whole purpose of this research project is to be able to exploit multimodal inputs even when not all sources are always available. 

Unfortunately, these findings also uncover the limitations of the proposed multi-encoder framework: In this instance, representing the important situation of data originating from YouTube, vision-related content is inaccessible for about 15.5\% of all samples. For this case, the gains our proposed method is able to achieve are certainly worthwhile. If the amount of examples with visual assistance  would diminish further, the performance boosts would also inevitably lessen, up to a point where it may no longer be worth the extra effort. 

\subsubsection{Tri-encoder attention-based models}

Table~\ref{tab:tri} lists the scores obtained by models that use three encoders: One takes in spectral auditory maps, the two others accept OpenL3 and temporally coherent visual features respectively. Pretrained VGG16 embeddings are not investigated any further: As discussed at length in the previous sections, in the bi-encoder experimental configuration, these vectors have already been found to be unsuitable for the tasks at hand, particularly for fine-grained sound event detection.

The interpolation hyperparameters are decided using procedures similar to those employed in the previously described experiments. As a first option, their values are chosen by optimizing event-based F1 scores on the validation partition of the considered data set. We test the following possibilities during training as well as inference for the auditory information weight: 0.25, 0.5, 0.75 and 0.875. Alternatively, the learning weights linked to the acoustic inputs are sampled randomly from a uniform distribution ranging from 0.25 to 1. To limit the search space in both of the foregoing cases, we choose to split the remaining share equally between the encoders connected to the visual streams. 

When comparing results obtained by audio-only models (in Table~\ref{tab:unibi}) to those of bi-encoder systems also incorporating OpenL3 or temporally coherent visual embeddings (in Table~\ref{tab:unibi} and Table~\ref{tab:birandom}), we mostly find consistent and substantial improvements. This has been discussed at length in the previous section(s). When inspecting the scores produced by tri-encoder architectures utilizing all three sets of relevant features (Table~\ref{tab:tri}), we observe additional performance increases. However, these boosts are far less pronounced.

This makes sense as, in contrast to the initial switch from unimodal to bi-encoder models, we are not integrating a new modality into the systems, we are simply using more sets of (visual) features. It is very likely that much of the information captured in one series of vision-related embeddings is also present in the other, leading to diminishing returns in terms of performance when combining them. However, this experiment does demonstrate the flexibility of the proposed framework and how it could be employed when more than two groups of complementary features are available.

\section{Conclusion}
\label{sect:con}

We proposed a dynamic multi-encoder approach to deal with the problem of missing values in the context of multimodal sound recognition. This particular situation frequently occurs since many pertinent data sets stem from YouTube, which provides a noteworthy opportunity but also poses a serious challenge: Visual inputs can be taken advantage of to enhance audio tagging and sound event detection models, which traditionally only employ acoustic information. However, vision-related features may not be accessible for all data points due to a variety of availability issues. We applied the aforementioned method to state-of-the-art attention-based neural network architectures. 

We performed experiments using the data set associated with task 4 of the DCASE 2020 challenge and verified that the presented framework can lead to noteworthy performance boosts in a selection of different evaluation settings. We thoroughly investigated the outcomes of said trials and analyzed some properties and limitations of the introduced technique. Among other things, we showed that improvements were contingent on a good choice of (pretrained) visual features to be used in conjunction with spectral auditory maps and we demonstrated that the proposed method even holds up when all vision-related inputs are ignored.

The proposed framework is naturally flexible, and consequentially, there are some compelling possibilities for future research involving this principle. Firstly, further attention could be directed to the way encoder interpolation hyperparameters are set during training and inference: In this work, we started with fixed weights, optimized on held-out validation data. Afterwards, we showed that randomizing these values during training can lead to similar results and significantly less tuning. It could be interesting to investigate more complex schemes, such as data-dependent weighting. Secondly, other sets of features could be explored. Here, we stuck to utilizing pretrained visual features on top of rudimentary spectral auditory maps, but pretrained auditory features might be worthwhile as well. Lastly, this technique is in no way tied to sound recognition, and it could easily be applied to other research tasks which also encounter missing value problems. 



\begin{backmatter}

\section*{Abbreviations} 
DCASE: Detection and Classification of
Acoustic Scenes and Events; ReLU: Rectified linear unit; CBF1: Clip-based F1 score; SBF1: Segment-based F1 score; EBF1: Event-based F1 score; PSDS: Polyphonic sound event detection score

\end{backmatter}

\section*{Declarations}

\begin{backmatter}

\section*{Availability of data and materials}
The data set analyzed during the current study is available in the DCASE 2020 task 4 repository, \href{http://dcase.community/challenge2020/task-sound-event-detection-and-separation-in-domestic-environments}{http://dcase.community/challenge2020/task-sound-event-detection-and-separation-in-domestic-environments}.

\section*{Competing interests}
The authors declare that they have no competing interests.

\section*{Funding}
Not applicable.

\section*{Authors' contributions}
Wim Boes conducted the research and performed the experiments. Hugo Van hamme guided and supervised the project. All authors read and approved the final manuscript.

\section*{Acknowledgements}
This work was supported by a PhD Fellowship of Research Foundation Flanders (FWO-Vlaanderen) and the Flemish Government under ``Onderzoeksprogramma AI Vlaanderen''.


\bibliographystyle{bmc-mathphys} 
\bibliography{bmc_article}      

\end{backmatter}
\end{document}